# 3D manipulation with scanning near field optical nanotweezers

J. Berthelot [1], S. S. Aćimović [1], M. L. Juan [2,3], M. P. Kreuzer [1], J. Renger [1] and R. Quidant [1, 4, ★]

[1] ICFO - Institut de Ciencies Fotoniques, Mediterranean Technology Park, 08860 Castelldefels (Barcelona), Spain
[2] *Department of Physics & Astronomy, Macquarie University, Sydney, NSW 2109, Australia*
[3] *ARC Centre for Engineered Quantum Systems, Macquarie University, Sydney, NSW 2109, Australia*
[4] *ICREA – Institució Catalana de Recerca i Estudis Avançats, 08010 Barcelona, Spain*

**Recent advances in Nanotechnologies have prompted the need for tools to accurately and non-invasively manipulate individual nano-objects [1]. Among possible strategies, optical forces have been foreseen to provide researchers with nano-optical tweezers capable to trap a specimen and move it in 3D [2-4]. In practice though, the combination of weak optical forces involved and photothermal issues have thus far prevented their experimental realization. Here, we demonstrate first 3D optical manipulation of single 50 nm dielectric objects with near field nanotweezers. The nano-optical trap is built by engineering a bowtie plasmonic aperture at the extremity of a tapered metal-coated optical fiber. Both the trapping operation and monitoring are performed through the optical fiber making these nanotweezers totally autonomous and free of bulky optical elements. The achieved trapping performances allow for the trapped specimen to be moved over tens of micrometers during several minutes with very low in-trap intensities. This novel non-invasive approach is foreseen to open new horizons in nanosciences by offering an unprecedented level of control of nano-sized objects including heat-sensitive bio-specimens.**

Further understanding physical, chemical or biological sciences often requires fractionating complex and macroscopic mechanisms into simpler elementary ones occurring on the nanometer scale. To this aim, researchers have put lots of efforts in developing techniques for monitoring as well as controllably and non-invasively manipulating elementary units of matter down to the single atom/molecule level. Among these techniques, scanning probe microscopy (SPM) has played a key role. Originally developed for imaging purposes, Scanning Tunnelling Microscopy (STM) became a powerful way to manipulate individual atoms adsorbed at a surface [5]. Slightly later, the Atomic Force Microscope (AFM) followed a similar route and resulted useful to pick up and move slightly larger amounts of matter, typically in the range of tens of nm, such as nanocrystals, nanoparticles or carbon nanotubes [6,7]. Interestingly, the optical version of scanning probe microscopes, namely Near Field Scanning Optical Microscopy (NSOM) [8,9], was originally foreseen to also have nano-manipulation capability. Back in the 90's, several theoretical studies predicted the strong optical concentration produced at the tip of a sharply elongated metallic probe, should create optical forces strong enough to stably trap dielectric objects as small as a few nanometers [2-4,10-12]. Despite these predictions, that has never been reported experimentally, mainly prevented by photothermal effects [9,13]. While this configuration offers strong optical gradients, the required level of local field intensity within the trap, larger than $10^{12}$ W/m$^2$ when accounting for the large intensity enhancement at the tip (*e. g.* 3000 [2]), is susceptible to damage either the specimen to trap [14] or the tip itself [15-18]. When operating in liquid, absorption within the metal is also responsible for heat-induced fluid dynamics [19] or bubble formation [20] that may further jeopardize trapping. Herein, we report on the first realization of optical trapping and manipulation of an individual nano-object at the extremity of a NSOM probe. Our nano-optical tweezers is formed at the extremity of a metal-coated tapered optical fiber patterned with a bowtie nano-aperture (BNA). We demonstrate stable optical trapping and accurate 3D manipulation of a 50 nm polystyrene (PS) bead in water with local intensities within the trap as small as $10^9$ W/m$^2$. Such level of intensity sits well below what would be required for conventional optical tweezers (typically from $10^{11}$ to $10^{12}$ W/m$^2$) and are compatible with heat-sensitive objects like bio-specimens.

Our present work capitalizes on the latest advances in near field optical trapping [21-23] based on the so-called Self-Induced Back-Action (SIBA) mechanism that dramatically relaxes the requirements on the local optical intensity and thus minimizes photothermal issues [24]. In SIBA trapping, the resonant optical nanostructure is designed such that its optical properties (resonance spectrum, local field distribution and intensity) significantly depend on the presence of the specimen. For a trapping laser slightly red-shifted with respect to the central resonance wavelength of the nanostructure, the trap becomes stiffer when the specimen tends to escape as the result of the induced resonance shift. In other words, the trapped specimen plays an active role in the trapping mechanism in a way that the required average local field intensity is weaker by orders of magnitude when compared with conventional trapping. Interestingly, the trap reconfiguration does not require any active monitoring of the specimen as it is automatically synchronized with its dynamics [25-27].

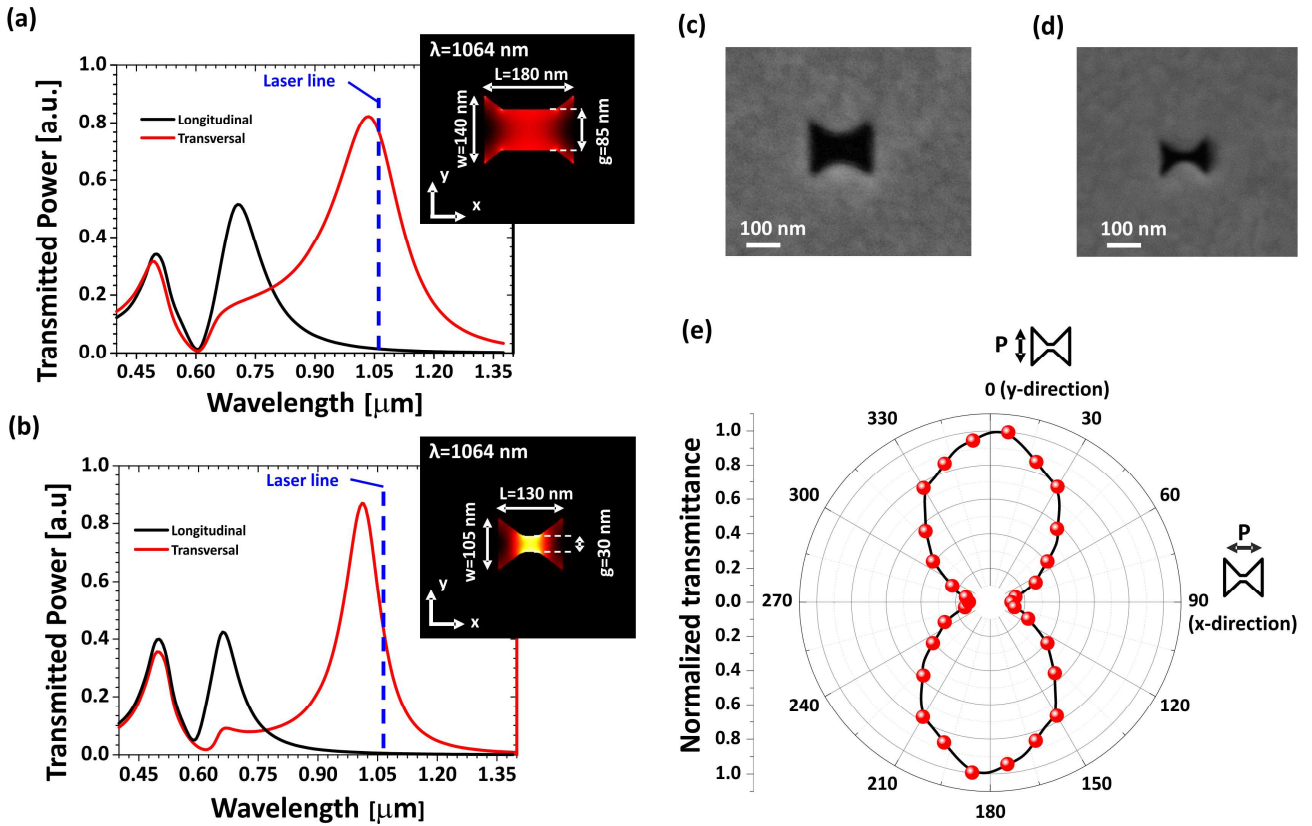

**Figure 1: Optical properties of Bowtie Nano Apertures (BNA).** (a-b) Calculated transmission spectra and near field intensity map at 1064 nm under transverse incident polarization for the two fabricated BNAs presented in SEM micrographs (c) and (d), respectively. (e) Evolution of the experimental transmission at 1064 nm as a function of the incident polarization for a 30 nm gap BNA.

In practice, implementing SIBA trapping at the extremity of a tapered optical fiber first requires identifying a geometry of the SIBA trap enabling extended trapping times under low laser intensity, to prevent any photothermal damage at the fiber extremity. Typical damage threshold for such probes sits in the range of $10^{10}$ W/m² concentrated at the tip apex [18]. In order to remain below this threshold, we focused our attention on the so-called BNA design [28-30]. This geometry combines high collection cross-section and transmission with strong mode confinement under transversal polarization, making of it a very good candidate for SIBA trapping. BNA support two types of resonances: a Fabry-Pérot-like resonance that mainly depends on the film thickness along with two plasmonic resonances that depend on the geometry of the aperture [29]. Extensive 3D numerical simulations based on COMSOL were performed to identify the most suitable design to achieve SIBA trapping at 1064 nm on a 100 nm-thick gold covered with water. Parameters were chosen such that the transverse plasmonic mode (confined within the gap region) is slightly blue-detuned with respect to the trapping wavelength. This condition is fulfilled for gap sizes between 30 nm and 85 nm by adjusting both the length and width of the antenna. Based on these simulations, BNAs were first fabricated in a planar geometry to evaluate and optimize their trapping performance. Figures 1.(c) and (d) show the SEM micrographs of two fabricated apertures, with 85 nm and 30 nm gaps, respectively. The transmission spectra and near field maps at 1064nm displayed in Fig. 1(a-b) were calculated from the experimental geometrical parameters extracted from Fig.1 (c-d). As expected the transverse mode confinement increases for decreasing gap sizes. In agreement with prior studies, the enhancement of the local electric field intensity at the center of the gap ranges from below 100 for 30 nm gap to below 10 for 85 nm gap [31]. In order to probe the transverse mode of the fabricated BNA, we also measured the evolution of the transmission through the aperture at 1064 nm as a function of the incident polarization. In good agreement with the simulations, the polar transmission plot of a 30 nm gap BNA (Fig. 1(e)) features a maximum when the incident electric field aligns across the gap (y-axis).

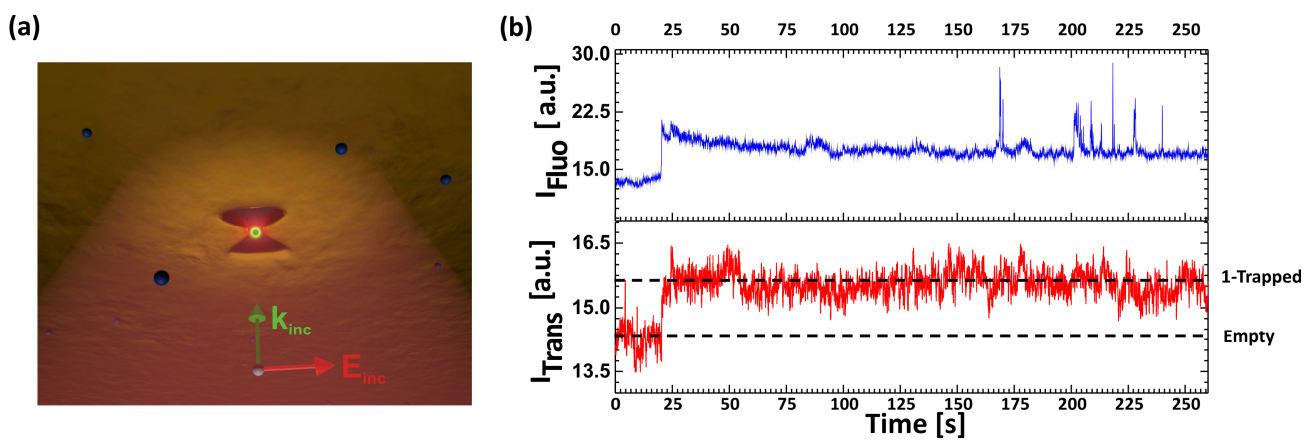

**Figure 2: SIBA trapping of a single 20 nm PS bead in a planar geometry.** (a) Schematic of the experimental configuration. A 30 nm gap BNA is illuminated from the waterside with a 1064 nm laser beam, linearly polarized along the y-axis, and slightly focused with a 40x (0.65 NA) objective lens. (b) Experimental time traces showing the transmission through the BNA at 1064 nm (red curved) and the fluorescence from the trapped bead (blue curve). The increase in both transmission and fluorescence corresponds to the trapping of a single 20 nm PS bead.

Figure 2 illustrates the trapping performance of a BNA with a 30 nm gap. The aperture was exposed to a diluted aqueous solution (0.05% w/v) of 20 nm fluorescent polystyrene (PS) beads (absorption@532 nm / emission@612 nm) containing 1% concentration of sodium dodecylsulphate (SDS) solution to prevent aggregation. Figure 2(a) gives a schematic of the experimental configuration. The 1064 nm-trapping laser was linearly polarized along the y-axis before being slightly focused (from the water side) onto the BNA with a 40x objective lens (0.65 NA), producing an illumination spot of about 2 µm. A 532 nm laser was added to the same optical path to simultaneously excite the bead fluorescence. The transmission through the BNA at 1064 nm and the fluorescence from the bead were monitored over time by two independent photodetectors (see method section). A typical portion of the experimental time traces demonstrating optical trapping of a single 20 nm bead for an irradiance of 1,27 x$10^9$ W/m² is presented in Fig. 2(b). Trapping is monitored by the increase of the transmission signal resulting from the local increment of refractive index induced by the presence of the particle [24–26]. Statistical analysis of the data enables identifying two transmission levels corresponding to an empty trap and trapping of a single bead. This interpretation is further confirmed by the fluorescence time trace that displays an increase exactly coinciding with the increased level of transmission. Trapping times longer than 30 min were achieved under these illumination conditions. We also systematically checked that the trapped nanoparticle did not stick to the antenna and was released when switching off the trapping laser (Fig. SI1 of the SI). To test the robustness of SIBA trapping with the BNA geometry, we also performed additional experiments using larger gaps of 85 nm. Despite the weaker confinement of the mode (translating in a weaker effect of the trapped specimen on the BNA resonance), both 50 nm and 20 nm PS beads where successfully trapped over several minutes using incident powers of 2 and 5 mW (0,63 ×$10^9$ W/m² and 1,59 ×$10^9$ W/m²), respectively.

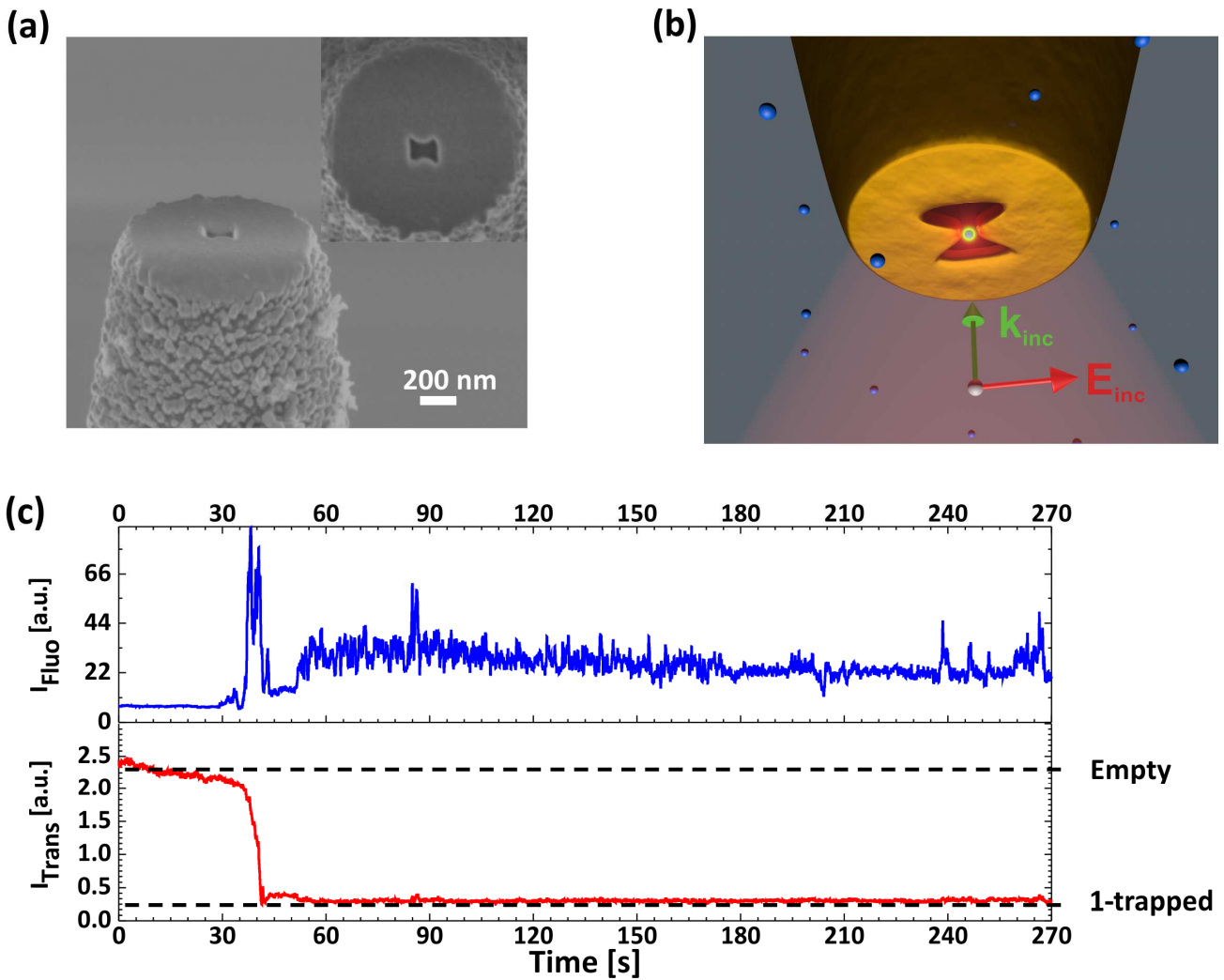

**Figure 3: SIBA trapping at the extremity of a patterned tapered fiber under external illumination.** (a) SEM images of an 85 nm gap BNA patterned at the extremity of a tapered optical fiber. (b) Schematic of the experimental configuration: the antenna is illuminated from the water side with a 1064 nm laser beam focused through a 40x (0.65 NA) objective. The incident polarization is aligned along the BNA gap to excite its transverse mode. (c) Portion of the time traces of the 1064 nm signal detected by the fiber (red) and the fluorescence from the bead (blue) detected by the objective lens. A 50 nm PS bead enters the aperture area around t=35 s and gets stably trapped after t=40 s.

At this stage, optimized in-plane BNA trapping was implemented at the extremity of a scanning tapered metallized optical fiber to achieve nano-optical tweezers capable to manipulate the trapped specimen in three dimensions. The fabrication process used to prepare the tapered fibers was adapted from [18]. Briefly, a single mode optical fiber was tapered by laser pulling and coated with 200 nm aluminium to prevent light leakage through the sides of the cone. FIB milling was used to cut the tip extremity to obtain a flat facet that was subsequently metallized with 100 nm of gold. Finally, the BNA was milled by FIB at the centre of the 1-μm diameter gold platform.

An example of a fabricated tweezers is shown in the SEM micrograph of Figure 3(a). It is worth mentioning that, with the FIB we used, the fabrication of the 30 nm gap BNAs required exposure parameters that were not compatible with fully preserving the integrity of the fiber output facet. For this reason, in the following, we limit ourselves to gaps of 85 nm. For the trapping experiments, the fabricated fibers were mounted on a 3D piezoelectric scanner and introduced into a modified fluidic chamber (see Methods section) containing solution of 50 nm PS bead (same parameters as in the in-plane experiments). We tested two trapping schemes based on different illumination conditions; either through an objective lens or through the fiber itself.

In the first configuration schematically presented in Fig. 3(b), the 1064 nm laser is focused through the 40x objective (0.65 NA) and centred on the BNA. The transmitted light is collected through the fiber and sent to a silicon photodetector. The fluorescence from the bead is collected back through the same objective lens before being focused on an avalanche photodiode using a confocal detection. Figure 3(c), shows a portion of typical time traces demonstrating the trapping of a single 50 nm bead using an incident power of 4 mW (1,27 x10$^9$ W/m²). In this configuration, unlike what was observed in the planar sample, the trapping event is associated to a decrease in the transmitted signal. Such decrease is attributed to the detection through the fiber. Because of its sub-wavelength size, the nanobead scatters light over a wide range of k-vectors, thus decreasing the amount of light coupled into the fiber mode. This interpretation is consistent with the simultaneous increase of the fluorescence signal confirming the presence of the nanobead within the BNA. Like for the in-plane configuration we verified the particle is released when switching off the trapping laser (Fig. SI2 of the SI). Although stable trapping was achieved for more than 3 minutes, this configuration bears some issues. First of all, the trap position is fixed and does not take advantage of the tip mobility. Secondly, because the tapered fiber and the collection objective are physically independent, vibrations and small drifts in their relative position produce additional noise making the monitoring and analysis of the trapping events more difficult.

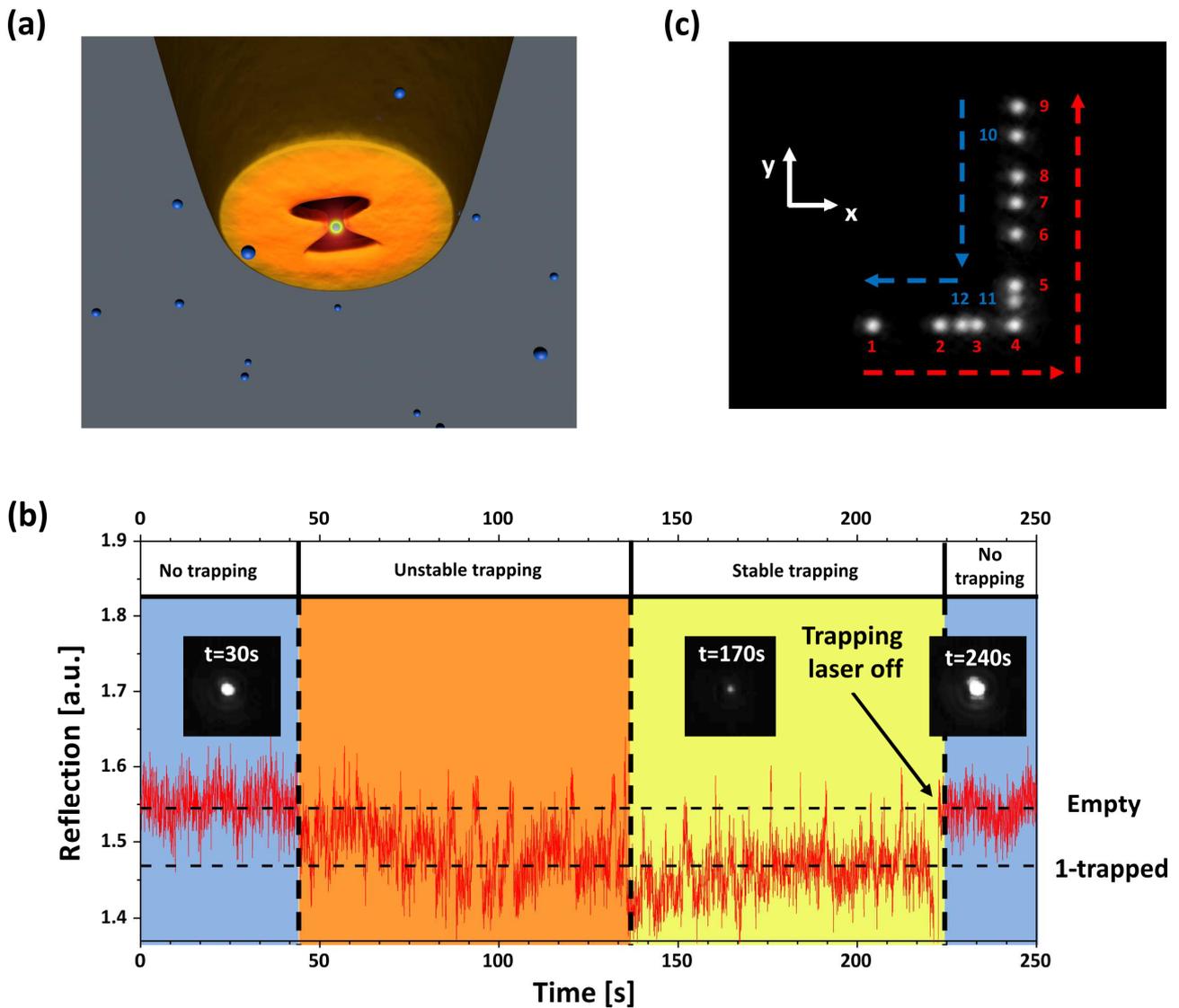

**Figure 4: 3D manipulation of a single 50 nm PS bead.** (a) Schematic of the experimental configuration. The 1064 nm trapping laser is directly coupled into the fiber to excite the transverse mode of the BNA. (b) Experimental time trace of the reflected 1064 nm signal showing trapping of a single 50 nm PS bead. Insets: optical images of the fiber transmission spot at different times. (c) Composite image reproducing the displacement of the trapped object. This displacement takes place during the time period t=180-210 s of time trace (b). See also the video in the Supplementary Information.

To overcome these issues, we switch to the second configuration where both illumination and collection are performed through the fiber. For this purpose we implemented a (90/10) fiber coupler combined with a polarization maintaining (PM) fiber designed for 1064 nm. The polarization of the incoming 1064 nm laser (10% arm) is controlled at the entrance of the PM fiber and set to maximize the transmission through the BNA. The 1064 nm light reflected by the tapered end face of the fiber is then collected by the 90% arm and focused on a silicon photodetector. A schematic of the experimental configuration is given in Figure 4(a). With this configuration both excitation and detection can be made by the only use of the fiber. Additionally, the transmission collected by a 40x objective (0.65 NA) is imaged on a charge coupled device (CCD) camera for monitoring purposes. A typical time trace of the reflected signal through the fiber is plotted in Figure 4(b). In order to estimate the incident intensity reaching the BNA plane, power measurements were systematically performed on aluminized tapered fibers prior to the deposition of the gold layer. Powers ranging from 350 to 675 µW, depending on the taper angle, were measured at the output of the fiber output facet. The corresponding local intensities ranging from $4 \times 10^9$ W/m$^2$ to $8 \times 10^9$ W/m$^2$ (accounting for the 10-fold theoretical intensity enhancement within the antenna gap) sit below the fiber damage threshold [18]. Similarly to the previous trapping experiment, we identify in Fig. 4(b) two different levels in the reflected 1064 nm signal corresponding to the empty trap (higher level) and trapping of a single nanobead (lower level) (see Fig. SI3 of the SI). Trapping also corresponds to a decrease in the intensity of the transmitted signal recorded by the CCD camera. A closer look to the time trace of Figure 4(b) enables us to distinguish, three successive regimes: no trapping (blue area), unstable trapping (orange area) and stable trapping (red area). During the unstable trapping regime (40 s < t < 135 s), the nanobead moves within the BNA without reaching the equilibrium position in the gap region, leading to large fluctuations in both the reflected and the transmitted signals. As this behaviour was not observed in the previous configuration, this suggests that the illumination through the objective helps to push the particle towards the equilibrium position. Around t=140 s, both signals become more steady, entering in a stable trapping regime. Once the stable trapping regime was reached, the fiber was raster scanned in all 3 directions of space (see video Supplementary Information). The scan trajectory is shown in Fig. 4(c) in which we have superimposed different CCD images from different time instances. The displacement of the tip was around 15 µm in the plane (x,y) and around 5 µm in the z direction. Finally, we verified that the trapped nanobead did not stick to the BNA and can be released by blocking the trapping beam (t=215 s).

In conclusion, we have developed the first nano-optical tweezers capable of 3D manipulating with nanometer accuracy sub-100 nm dielectric objects over large ranges. We envision this approach could become a novel enabling nanotool that could potentially open new opportunities in various fields of science. In the context of biology it may enable non-invasive manipulation of individual nano-units such as viruses or large proteins [32]. It may also benefit the field of material science with the possibility to isolate, manipulate and controllably arrange solid-state nano-objects such as nanocrystals.

**Acknowledgments:**

This work was partially supported by the Spanish Ministry of Sciences (grants FIS2010– 14834, MAT2010-14885 and CSD2007–046-NanoLight.es), the European Community's Seventh Framework Program under grant ERC-Plasmolight (no. 259196) and Fundació privada CELLEX. The authors thank M. Mivelle and M. García-Parajo for fruitful discussions.

**Methods**:

Microfluidic chamber for fiber trapping:

The fluidic chamber used for fiber trapping was made of Polydimethylsiloxane (PDMS) with curing agent (5:1 in volume). With this mixture we obtained a PDMS piece of 20 mm x 20 mm x 5 mm. The chamber volume is made by removing the inner part of this piece. A thin membrane made with a solution PDMS/curing agent 20:1 in volume, is fixed on the top to prevent liquid evaporation. A hole (2 mm in diameter) was made on the centre of the membrane, allowing the introduction of the structured fiber. The complete PDMS part is chemically bonded (12 hours at 80°C in the oven) to a glass substrate of 170 µm thick. The chamber was then completely filled with the solution of nanobeads through the use of an inlet.

Optical Setup:

We used a homemade inverted microscope. The trapping laser is a CW 1064 nm Nd-YAG laser. Its bean is extended and collimated to an 8 mm diameter beam before being focused on the sample plane with a dry 40x microscope objective (0.65 NA). A visible diode laser at 532 nm is focused on the sample at the same position than the trapping laser to excite the bead fluorescence and. The same objective is used for the fluorescence excitation and its collection. The collected fluorescence is separated from the trapping laser beam with a dichroic mirror (reflecting 532 nm and 1064nm and transmitting others) and band pass filter (580 nm to 640 nm), and then focused onto an avalanche photodiode in a confocal detection mode. The transmission of the trapping laser through the BNA is collected with a 10x dry objective (0.22 NA) and send to a photodiode. The polarization of the 1064 nm laser is controlled by a polarizer and a half wave plate. The incoming laser power is limited to maximum 10 mW in the sample plane. The acquisition signals are done by the use of a Labview program at a sampling rate of 1 kHz

**References**

[1] Maragò, O. M., Jones, P. H., Gucciardi, P. G., Volpe, G., & Ferrari, A. C. Optical trapping and manipulation of nanostructures. *Nature Nano*, **8**(11), 807–19 (2013).

[2] Novotny L., Bian R. & Xie, X. Theory of Nanometric Optical Tweezers. *Phys. Rev. Lett.* **79**, 645 (1997).

[3] Martin, O.J.F. & Girard, C. Controlling and tuning strong optical field gradients at a local probe microscope tip apex. *Appl. Phys. Lett.* **70**, 705 (1997).

[4] Chaumet, P., Rahmani, A. & Nieto-Vesperinas, M. Optical Trapping and Manipulation of Nano-objects with an Apertureless Probe. *Phys. Rev. Lett.* **88**, 123601 (2002).

[5] Hla, S.-W. Scanning tunneling microscopy single atom/molecule manipulation and its application to nanoscience and technology. *J. Vac. Sci. Technol. B* **23**(4), 1351 (2005).

[6] Kim, S., Shafiei, F., Ratchford, D. & Li, X. Controlled AFM manipulation of small nanoparticles and assembly of hybrid nanostructures. *Nanotechnology* **22**(11), 115301 (2011).

[7] Decossas, S., Patrone, L., Bonnot, a. M., Comin, F., Derivaz, M., Barski, A. & Chevrier, J. Nanomanipulation by atomic force microscopy of carbon nanotubes on a nanostructured surface. *Surface Science* **543**(1-3), 57–62 (2003).

[8] Betzig, E., Lewis, A., Harootunian, A., Isaacson, M. & and Kratschmer, E. Near Field Scanning Optical Microscopy (NSOM): Development and Biophysical Applications. *Biophysical J.* **49**(1), 269–279 (1986).

[9] Hecht, B. *et al*. Scanning near-field optical microscopy with aperture probes: Fundamentals and applications. *J. Chem. Phys.* **112**(18), 7761 (2000).

[10] H'dhili, F., Bachelot, R., Lerondel, G., Barchiesi, D. & Royer, R. Near-field optics: Direct observation of the field enhancement below an apertureless probe using a photosensitive polymer. *Appl. Phys. Lett.* **79**, 4019 (2001).

[11] Chaumet, P., Rahmani, A., & Nieto-Vesperinas, M. Photonic force spectroscopy on metallic and absorbing nanoparticles. *Phys. Rev. B* **71**, 045425 (2005).

[12] Okamoto, K., & Kawata, S. Radiation Force Exerted on Subwavelength Particles near a Nanoaperture. *Phys. Rev. Lett.* 83, 4534–4537 (1999).

[13] L. Novotny, private communication.

[14] Ashkin, A., Dziedzic, J., Bjorkholm, J. & Chu, S. Observation of a single-beam gradient-force optical trap for dielectric particles in air. *Opt. Lett.* **22**(11), 816–8 (1986).

[15] Xin, H., Xu, R. & Li, B. Optical trapping, driving, and arrangement of particles using a tapered fibre probe. *Scientific Reports* **2**, 818 (2012).

[16] Liberale, C. *et al.* Miniaturized all-fibre probe for three-dimensional optical trapping and manipulation. *Nature Photon.* **1**(12), 723–727 (2007).

[17] Liu, Z., Guo, C., Yang, J., & Yuan, L. Tapered fiber optical tweezers for microscopic particle trapping: fabrication and application. *Optics Express* **14**(25), 12510–6 (2006).

[18] Neumann, L. *et al*. Extraordinary Optical Transmission Brightens Near-Field Fiber Probe. *Nano Lett.* **11**(2), 355–60 (2011).

[19] Donner, J., Baffou, G., McCloskey, D. & Quidant, R. Plasmon-Assisted Optofluidics. *ACS Nano* **5**(7), 5457–5462 (2011).

[20] Fang, Z., Zhen, Y.-R., Neumann, O., Polman, A., García de Abajo, F. J., Nordlander, P., & Halas, N. J. Evolution of light-induced vapor generation at a liquid-immersed metallic nanoparticle. *Nano Lett.*, **13**(4), 1736–42 (2013).

[21] Righini, M. *et al*. Nano-optical trapping of Rayleigh particles and Escherichia coli bacteria with resonant optical antennas. *Nano Lett.* **9**, 3387-3391 (2009).

[22] Grigorenko, A. N., Roberts, N. W., Dickinson, M. R. & Zhang, Y. Nanometric optical tweezers based on nanostructured substrates. *Nature Photon.* **2**, 365–370 (2008).

[23] Juan, M. L., Righini, M. & Quidant, R. Plasmon nano-optical tweezers. *Nature Photon.* **5**(6), 349–356 (2011).

[24] Juan, M. L., Gordon, R., Pang, Y., Eftekhari, F. & Quidant, R. Self-induced back-action optical trapping of dielectric nanoparticles. *Nature Physics* **5**(12), 915–919 (2009).

[25] Pang, Y. & Gordon, R. Optical Trapping of 12 nm Dielectric Spheres Using Double-Nanoholes in a Gold Film. *Nano Lett.* **11**(9), 3763–3767 (2011).

[26] Chen, C. *at al*. Enhanced Optical Trapping and Arrangement of Nano-Objects in a Plasmonic Nanocavity. *Nano Lett.* **12**(1), 125-132(2012).

[27] Descharmes, N., Dharanipathy, U. P., Diao, Z., Tonin, M. & Houdré, R. Observation of Backaction and Self-Induced Trapping in a Planar Hollow Photonic Crystal Cavity. *Phys. Rev. Lett.* **110**(12), 123601 (2013).

[28] Kinzel, E. C. & Xu, X. Extraordinary infrared transmission through a periodic bowtie aperture array. *Opt. Lett.* **35**(7), 992–4 (2010).

[29] Guo, H. *et al*. Optical resonances of bowtie slot antennas and their geometry and material dependence. *Optics Express* **16**(11), 7756–66 (2008).

[30] Jin, E. X. & Xu, X. Plasmonic effects in near-field optical transmission enhancement through a single bowtie-shaped aperture. *Applied Physics B* **84**(1-2), 3–9 (2006).

[31] Mivelle, M., van Zanten, T. S., Neumann, L., van Hulst, N. F., & Garcia-Parajo, M. F. Ultrabright bowtie nanoaperture antenna probes studied by single molecule fluorescence. *Nano Lett.*, **12**(11), 5972–8 (2012)

[32] Pang, Y. and Gordon, R, Optical Trapping of a Single Protein, *Nano Lett*. **12**, 402–406 (2012)

# Supplementary information for "3D manipulation with scanning near field optical nanotweezers"

J. Berthelot [1], S. S. Aćimović [1], M. L. Juan [2,3], M. P. Kreuzer [1], J. Renger [1] and R. Quidant [1, 4, ★]

[1] ICFO - Institut de Ciencies Fotoniques, Mediterranean Technology Park, 08860 Castelldefels (Barcelona), Spain
[2] Department of Physics & Astronomy, Macquarie University, Sydney, NSW 2109, Australia
[3] ARC Centre for Engineered Quantum Systems, Macquarie University, Sydney, NSW 2109, Australia
[4] ICREA – Institució Catalana de Recerca i Estudis Avançats, 08010 Barcelona, Spain

### *Trapping vs. adsorption (planar sample)*

For each trapping experiment, we verified that the particle does not stick to the antenna and was indeed maintained in the trap by the optical forces. Initially the particle was stably trapped for several hundreds of seconds. In the following step the trapping laser was switched off to release the particle and we then checked that the transmission reverted back to its initial value i.e. before any particle was trapped. Indeed, in case the particle would be sticking to the antenna, no change in the transmission would be noted once the trapping laser had been switched off.

Figure SI1 presents the trapping of a single 20 nm PS bead via both the transmission through the antenna and the associated fluorescence of the bead. The particle is clearly trapped after t=15 s, the trapping laser is then switched off at t=336 s for 25 s. During this time, the fluorescence is still excited and shows important fluctuations. The latter ones have to be attributed to the fact that particles move across the detection volume.

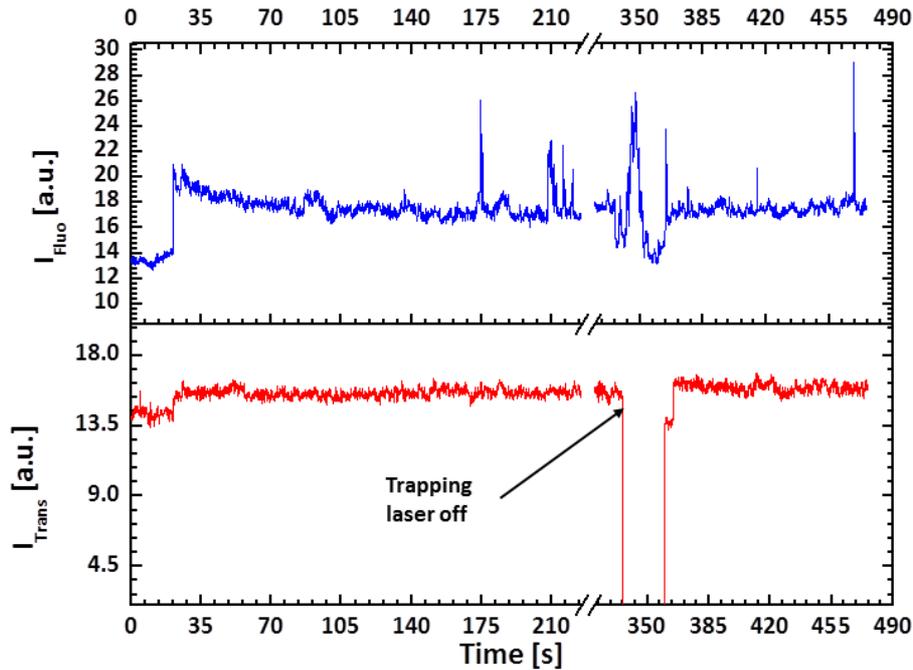

**Figure SI1 : Experimental time traces showing trapping and release of a 20 nm PS particle with an in-plane Bowtie nanoaperture:** Transmission (in red) and associated fluorescence (in blue). Around t=340 s the trapping laser is switched off while maintaining the green excitation laser on. The trapping laser is switched back on at t=360 s whereby a new particle gets trapped at t=375 s.

At around t=360 s the laser is switched back on. Both the transmission and the fluorescence recovered their initial levels, i.e. before any particle was trapped. Since the free moving particle did not travel far away from the trap while the trapping laser was off, the particle is trapped again briefly after switching the laser back on. The clear possibility to release the particle by simply switching off the trapping laser proves the particle was indeed trapped and not adsorbed.

*Trapping vs. adsorption (fiber configuration under external illumination)*

In a similar manner as in the case of a planar sample we checked that the particle was not adsorbed at the surface when using the fiber configuration under external illumination. For this particular configuration the fiber was fixed, the fluorescence was then collected through a microscope objective aligned in order to collect the fluorescence around the antenna. Figure SI2 presents the fluorescence of the bead and transmission through the antenna for the trapping of a 50 nm PS bead. At approximately t=30 s, both transmission and fluorescence show the trapping of a single particle. The particle was kept in the trap until t=270 s when the trapping laser was switched off. Shortly after (20 s), the trapping laser was switched back on. After this interruption of the trapping laser, the transmission and the fluorescence both reverted to their initial levels (before any particle was trapped), clearly proving the particle was not adsorbed to the surface.

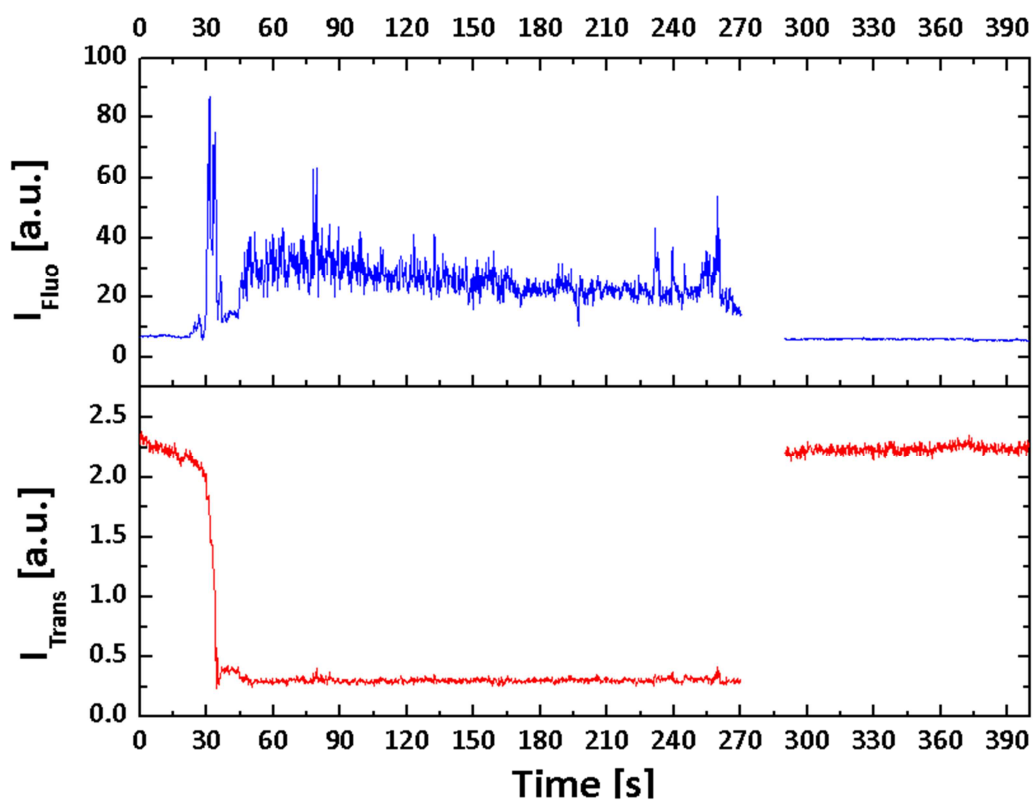

**Figure SI2 : Experimental time trace showing trapping and release of a single 50 nm trapped particle:** At t=270 s both the trapping laser and the green laser are switched off until t=290 s.

## *Trapping vs. adsorption (fiber configuration with illumination through the fiber)*

For this configuration, no objectives where used to collect the signal in order to obtain a self contained system consisting of only the fiber to both trap and monitor trapping by the change of the reflected light coupling backwards in the fiber mode. Consequently, no fluorescence signal was collected. Yet, we could verify the particle was not adsorbed during the experiment by examining the reflection from the antenna. Figure SI3 presents the signal back reflected from the antenna during the trapping and release of a 50 nm PS bead. After an unstable trapping phase (40 s < t < 135 s), the particle is stably trapped for more than 60 s until the trapping laser was switched off at t=215 s. Soon after the trapping laser was switched back on, and the signal clearly restored back to its initial level showing the particle was not adsorbed. To overcome the high noise level, the histogram was fitted using two Gaussian distributions related to the different states of the system: no particle trapped and one particle trapped. This double fit (in red) shows a very good agreement with the data points, clearly demonstrating one particle was trapped for over a minute, then released by simply switching off the trapping laser during a short time.

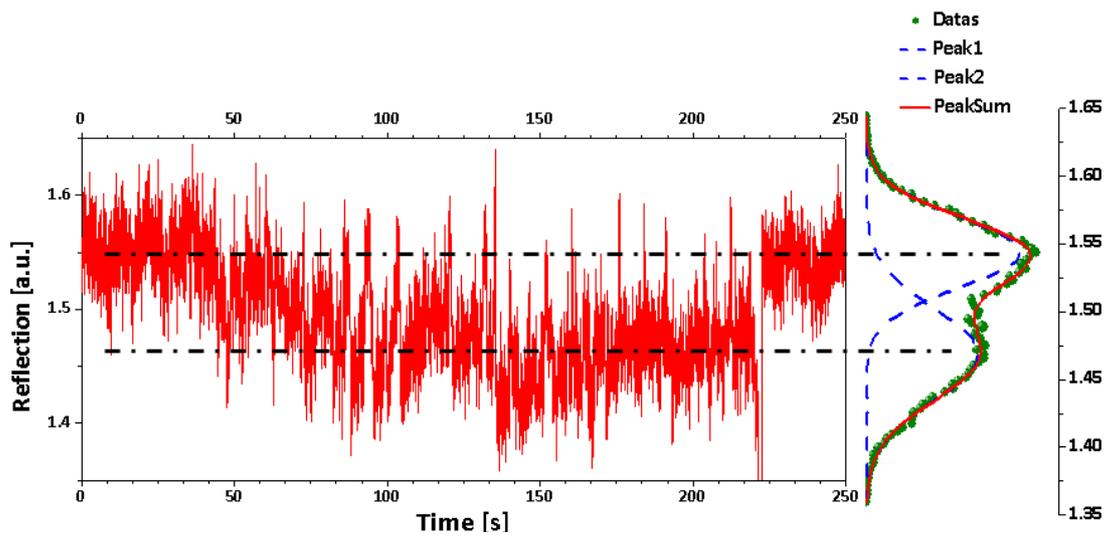

Figure SI3 : **Analysis of the time trace showing trapping of a 50 nm PS particle at the apex of the patterned fiber**: Histogram of the time trace (green dots). The dots are fitted with a sum of two Gaussian functions (red line). The centre of each Gaussian peak is located by a black dot-dashed line.